\documentclass[sigconf,nonacm]{acmart}

\renewcommand\footnotetextcopyrightpermission[1]{}
\AtBeginDocument{%
  }

\begin{document}

\title[Bibliographic Coupling as Weak Supervision for Scientific Embeddings]{Bibliographic Coupling as Weak Supervision for Adapting Scientific
  Embeddings: A Case Study on a Specialized Interdisciplinary Corpus}

\author{Roman Neruda}
\correspondingauthor
\email{roman@cs.cas.cz}
\orcid{0000-0003-2364-5357}
\author{Petra Vidnerová}
\email{petra@cs.cas.cz}
\orcid{0000-0003-3879-3459}
\author{Jiří Šejnoha}
\email{sejnoha@cs.cas.cz}
\affiliation{%
  \institution{Institute of Computer Science, \\The Czech Academy of Sciences}
  \city{Prague}
  \country{Czech Republic}
}

\author{Jakub Mihálik}
\email{mihalik@flu.cas.cz}
\author{Tomáš Marvan}
\email{marvan@flu.cas.cz}
\orcid{0000-0002-8487-5394}
\affiliation{%
  \institution{Institute of Philosophy, \\The Czech Academy of Sciences}
  \city{Prague}
  \country{Czech Republic}
}

\renewcommand{\shortauthors}{Neruda et al.}

\begin{abstract}
Scientific document embeddings such as SPECTER2 are optimized for retrieval over
large, heterogeneous corpora. On a small specialized corpus their similarity
distributions can become highly compressed, which limits their usefulness for
exploratory literature mapping. We study whether weakly supervised in-domain
adaptation, using only metadata already available in open citation databases,
changes this geometry. On ArtCon, a curated corpus of 704 publications on
artificial and machine consciousness, we adapt the SPECTER2 proximity model with
a lightweight LoRA adapter trained on positive pairs derived from bibliographic
coupling in OpenAlex, using no manually annotated similarity labels. We compare
the adapted embeddings with lexical (TF--IDF), citation-based (bibliographic
coupling) and generic semantic (SPECTER2) representations under thresholded,
fixed-density and symmetric $k$-nearest-neighbour graph constructions.
Adaptation substantially decompresses the similarity distribution (mean pairwise
cosine similarity 0.9104 to 0.8663, standard deviation 0.0259 to 0.0594) and
yields graphs with markedly higher Louvain modularity and better connectivity
than the base encoder under matched edge budgets and matched neighbourhood
budgets. The effect is stable across all three graph constructions and is
therefore not an artefact of a single sparsification choice. We report this as a
case study on one corpus, and we are explicit about what it does not establish:
the clustering indices used here are computed against community labels derived
from each representation's own similarity graph, so they measure the internal
geometric self-consistency of a representation rather than its agreement with an
external topical ground truth, and adaptation and evaluation use the same
documents. Evaluation against independent labels, on held-out documents, and
against geometric controls is the subject of ongoing work.
\end{abstract}

\keywords{scientific document embeddings, bibliographic coupling, weak
supervision, domain adaptation, literature mapping, science mapping, digital
libraries}

\maketitle

\section{Introduction}

The rapid growth of scientific literature has made exploratory analysis of scholarly corpora an increasingly important task for digital libraries. Beyond retrieving individual publications, researchers often seek to understand the thematic organization of a research field, identify relationships between research communities, and discover connections that are not immediately apparent from keyword search or citation links alone. These tasks require document representations that faithfully capture the semantic structure of a corpus and support meaningful similarity graphs, clustering, visualization, and interactive exploration.

Recent transformer-based models such as SPECTER~\cite{cohan-etal-2020-specter} and SPECTER2~\cite{singh-etal-2023-scirepeval} have substantially improved semantic representations of scientific documents by incorporating citation information during large-scale pre-training. These models achieve strong performance on retrieval benchmarks and have become widely used as general-purpose scientific embeddings. However, optimization for retrieval does not necessarily produce embedding spaces that are
well-suited for exploratory literature mapping. In particular, embeddings optimized for heterogeneous scientific collections may produce overly compressed similarity spaces in specialized domains, making it difficult to identify coherent thematic regions while preserving meaningful global connectivity.

Adapting scientific embedding models to a specialized research area is challenging because supervised fine-tuning typically requires manually annotated relevance judgments or domain-specific similarity labels, which are rarely available for emerging or interdisciplinary fields. A practical alternative is to exploit existing scholarly metadata as a source of weak supervision. Citation information is readily available through large-scale scholarly databases such as OpenAlex~\cite{priem2022openalex} and provides valuable signals about conceptual relationships between publications without requiring additional manual annotation.

In this work, we investigate whether bibliographic coupling can serve as such a weak supervision signal. Bibliographic coupling measures the extent to which two publications share references and has long been regarded as an indicator of topical relatedness~\cite{kessler1963}. Rather than using bibliographic coupling directly for visualization or clustering, we use it to adapt the embedding geometry itself. Specifically, we fine-tune the SPECTER2 proximity model using parameter-efficient LoRA adaptation, where strongly coupled document pairs define positive training examples. The resulting embeddings are then used for exploratory literature mapping.

We evaluate the proposed approach on ArtCon~\cite{artcon}, a curated corpus of 704 publications on artificial consciousness. The corpus spans philosophy of mind, artificial intelligence, cognitive science, neuroscience, psychology, and robotics, creating a challenging testbed in which multiple research traditions coexist while using only partially overlapping terminology and citation practices. Rather than evaluating the adapted embeddings solely through retrieval performance, we investigate whether they produce a representation space that is better suited for exploratory literature mapping. We compare lexical (TF--IDF), citation-based (bibliographic coupling), generic semantic (SPECTER2), and adapted semantic representations using graph connectivity, community structure, clustering quality, robustness across multiple graph-construction strategies, and qualitative neighbourhood analysis. To assess robustness, we evaluate these representations under thresholded similarity graphs, fixed-density graphs, and symmetric $k$-nearest-neighbour graphs.

The contributions of this paper are as follows:

\begin{itemize}
\item We describe a weakly supervised procedure for adapting scientific document embeddings using bibliographic coupling derived from citation metadata, requiring no manually annotated similarity labels and no full text.

\item We show that bibliographic coupling is a strong enough signal to substantially reshape the embedding geometry of a specialized corpus, rather than serving only as a bibliometric similarity measure applied at visualization time.

\item We evaluate the resulting structural changes under thresholded, fixed-density, and symmetric $k$-nearest-neighbour graph construction, showing that they are not artefacts of a particular graph-generation procedure.

\item We delimit what this evidence does and does not support, and identify the independent evaluations required to establish that the reorganized space is semantically, and not merely geometrically, better.
\end{itemize}

\paragraph{Scope.}
This paper reports a case study on a single curated corpus. The measures we use
characterize the geometry of each representation and the structure of the graphs
built from it; as discussed in Section~\ref{sec:method} and
Section~\ref{sec:limitations}, they are computed against labels derived from the
representation under evaluation and therefore cannot by themselves demonstrate
that one map of the literature is semantically more accurate than another. We
state our conclusions at that level throughout.

\section{Related Work}

Our work lies at the intersection of bibliometric science mapping, scientific document representation learning, and exploratory access to scholarly corpora.

The visualization of scientific literature as a network has a long tradition in bibliometrics. Classical measures of relatedness include bibliographic coupling \cite{kessler1963}, which counts the references two papers share, and co-citation \cite{small1973}, which counts how often later papers cite both together. Tools such as Bibliometrix \cite{bibliometrix}, Litmaps \cite{litmaps}, CiteSpace \cite{chen2006citespace}, and SciMAT \cite{cobo2012scimat} operationalize these ideas to produce static maps of research fields, while Gephi \cite{bastian2009gephi} provides general graph-layout support. Prior work has often found that combining textual and citation signals improves reconstruction of corpus structure \cite{boyack2010coclassification}. Our approach differs in using bibliographic coupling not at visualization time, but as a weak supervision signal for adapting a semantic encoder.

Transformer-based encoders have largely replaced bag-of-words and topic-model representations for scholarly corpora \cite{blei2003lda}. SPECTER introduced citation-informed contrastive learning for document-level scientific embeddings \cite{cohan-etal-2020-specter}, while SciNCL extended this idea with neighborhood-based contrastive sampling on citation graphs \cite{ostendorff-etal-2022-neighborhood}. SPECTER2 further unified this line of work through task-specific adapters trained on SciRepEval \cite{singh-etal-2023-scirepeval}. More recently, OpenMSD \cite{gao-etal-2024-openmsd} trains multilingual scientific-document similarity models with contrastive learning using direct citation, co-citation and bibliographic coupling as positive-pair signals, at web scale and with dedicated held-out test sets. Because SPECTER2 is trained predominantly on computer science and biomedicine, its embedding geometry may be suboptimal for smaller specialized corpora, which motivates the in-domain adaptation studied here. For clustering and exploratory organization of embedding spaces, recent work often combines UMAP \cite{mcinnes2020umapuniformmanifoldapproximation} with HDBSCAN \cite{campello2013hdbscan}, as in Top2Vec \cite{angelov2020top2vec} and BERTopic \cite{grootendorst2022bertopic}.

The components we combine are therefore established ones: a pre-trained scientific encoder, citation-derived training pairs, contrastive learning with in-batch negatives, and parameter-efficient fine-tuning. Our setting differs from the work above in scale and purpose rather than in mechanism. Instead of pre-training a general similarity model on a large citation corpus, we ask what happens when a frozen general-purpose encoder is adapted to one small, curated, interdisciplinary collection using only that collection's own coupling structure, and what can and cannot be concluded from the measures ordinarily used to evaluate such maps. The closest prior approaches, SciNCL and OpenMSD, are not evaluated as baselines here; we return to this in Section~\ref{sec:limitations}.

Parameter-efficient adaptation methods such as adapters \cite{houlsby2019adapters,pfeiffer-etal-2020-adapterhub} and LoRA \cite{hu2022lora} make it feasible to specialize large encoders with only a small number of trainable parameters, which is particularly attractive for small scholarly corpora where full fine-tuning would be expensive or unstable. In contrast to prior work that relies on large citation graphs or benchmark supervision, we derive a weak supervision signal directly from the corpus using bibliographic coupling computed from OpenAlex citation data \cite{priem2022openalex}.

Beyond static maps, user-facing tools such as Open Knowledge Maps \cite{kraker2016okmaps}, Connected Papers \cite{connectedpapers}, Litmaps \cite{litmaps}, and ResearchRabbit \cite{rabbit} support exploratory navigation of scientific literature. These systems typically rely on generic citation or text-similarity signals applied uniformly across domains. Our work is complementary: it targets a curated domain-specific corpus and asks whether in-domain adaptation can improve the representation space underlying exploratory literature mapping.

Existing work primarily evaluates scientific embeddings through retrieval performance or downstream NLP tasks. In contrast, we investigate whether domain adaptation changes the structural organization of embedding spaces used for exploratory literature mapping.

\section{Methodology}
\label{sec:method}
\subsection{Corpus and document representation}

Our experiments were conducted on ArtCon \cite{artcon}, a curated bibliography of work on artificial and machine consciousness containing 704 papers. The corpus spans several decades and brings together publications from philosophy of mind, artificial intelligence, cognitive science, and adjacent areas. This breadth makes it a useful test case for literature mapping: the corpus is topically coherent at a high level, but internally heterogeneous in terminology, methods, and argumentative style.

\begin{figure*}[t]
    \centering
    \includegraphics[width=.8\textwidth]{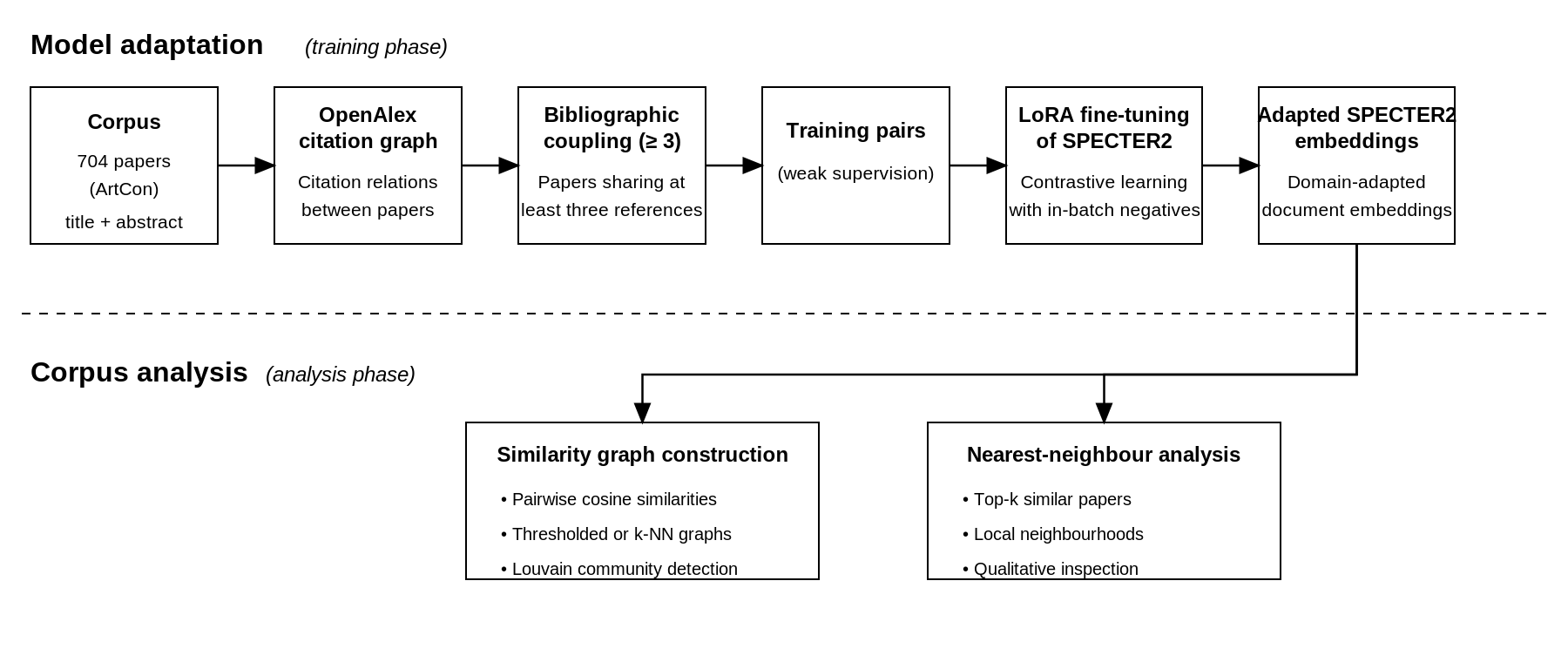}
\caption{
Overview of the proposed workflow. Bibliographic coupling extracted
from the OpenAlex citation metadata provides weak supervision for
parameter-efficient LoRA adaptation of the SPECTER2 encoder. The
resulting domain-adapted embeddings are used to construct similarity
graphs, detect communities, and analyse local semantic neighbourhoods
for exploratory literature mapping.
}
\Description{
Overview of the proposed methodology. A corpus of 704 scientific
papers is processed in two stages. During model adaptation,
bibliographic coupling computed from OpenAlex citation metadata is
used to generate positive training pairs for LoRA fine-tuning of the
SPECTER2 encoder. During corpus analysis, the adapted embeddings are
used to construct similarity graphs, identify Louvain communities,
and analyse nearest neighbours for exploratory literature mapping.
}
    \label{fig:methodology}
\end{figure*}

The proposed methodology consists of two stages. First, a general-purpose scientific embedding model is adapted to the target corpus using bibliographic coupling as a weak supervision signal. Second, the adapted embeddings are used for corpus analysis through similarity-graph construction and nearest-neighbour analysis. Figure~\ref{fig:methodology} summarizes the complete workflow.

For each paper, we constructed a document representation by concatenating the title and abstract. The same title-and-abstract representation was used for both the original and fine-tuned models. Because our goal is exploratory mapping of the corpus rather than full-text retrieval, we treat the title and abstract as a compact summary of each paper's thematic content. At inference time, each document is embedded into a dense vector space and similarities between papers are computed using cosine similarity.

\subsection{Base encoder}

As the base document encoder, we use SPECTER2 \cite{singh-etal-2023-scirepeval}, a transformer-based model for document-level scientific embeddings. SPECTER2 is pre-trained with citation-based supervision, so its representations are intended to reflect citation-informed relatedness as well as topical similarity; at inference time, however, it is given only the title and abstract of a paper, and no citation information about the papers being embedded is available to it. Its adapter-based architecture makes it suitable for parameter-efficient in-domain adaptation.

SPECTER2 builds on the SPECTER line of models, which exploit citation structure as supervision for representation learning \cite{cohan-etal-2020-specter}. Because its original training distribution is dominated by computer science and biomedicine, its embedding geometry may be suboptimal for a smaller specialized corpus such as ArtCon. Our method therefore starts from the pre-trained SPECTER2 proximity model and adapts it to the target domain.

\subsection{In-domain adaptation via bibliographic coupling}
\label{sec:adaptation}

To adapt the encoder to the artificial-consciousness literature, we fine-tuned SPECTER2 using bibliographic coupling as a weak supervision signal. Bibliographic coupling measures the extent to which two papers cite the same references; intuitively, papers that draw on overlapping prior work are likely to occupy related thematic positions in the corpus \cite{kessler1963}. Citation data were retrieved automatically through OpenAlex \cite{priem2022openalex}.

Positive training pairs were constructed from papers sharing at least three references. Negative examples were not mined explicitly: training used a multiple-negatives ranking objective \cite{henderson2017efficient}, in which, for each positive pair in a batch, the remaining documents of that batch serve as negatives. This formulation is well-suited to small corpora because it avoids explicit hard-negative mining while still encouraging the model to increase similarity for coupled papers and decrease similarity for uncoupled ones. It also means that the difficulty of the negatives is determined by batch composition rather than by a sampling policy.

Two thresholds appear in this paper and should not be confused. The positive-pair criterion used for \emph{training} is $\geq 3$ shared references, as recorded in Table~\ref{tab:lora-hyperparams}. The bibliographic-coupling \emph{baseline graph} reported in Section~\ref{sec:results} is constructed at $\geq 2$ shared references. Neither threshold was tuned, and the sensitivity of the adaptation to the positive-pair criterion is not examined in this paper; see Section~\ref{sec:limitations}.

Adaptation was performed with LoRA \cite{hu2022lora} on top of the pre-loaded SPECTER2 proximity adapter. The 110M-parameter base model was kept frozen, and only approximately 2M parameters were trained. We used title-and-abstract inputs truncated to 512 tokens, optimized with AdamW and a linear learning-rate schedule with warmup. The training configuration is summarized in Table~\ref{tab:lora-hyperparams}. No full text was used in either the original or fine-tuned condition. Training ran for a fixed number of epochs with no validation split and no early-stopping criterion, and all results below come from a single training run; we return to the consequences in Section~\ref{sec:limitations}.

\begin{table}[h]
  \centering
\caption{Hyperparameters used for LoRA adaptation of the SPECTER2 proximity
model. Only the LoRA parameters were optimized while the underlying
encoder remained frozen. Positive training pairs were constructed from
paper pairs sharing at least three references, and training was
performed using the Multiple Negatives Ranking Loss. The configuration
was selected to provide stable adaptation on the relatively small
ArtCon corpus.}
  \label{tab:lora-hyperparams}
  \begin{tabular}{lc}
    \hline
    \textbf{Hyperparameter} & \textbf{Value} \\
    \hline
    LoRA rank ($r$)         & 16 \\
    LoRA $\alpha$           & 32 \\
    LoRA dropout            & 0.1 \\
    Trainable parameters    & $\approx$2M / 110M \\
    Max sequence length     & 512 tokens \\
    Batch size              & 16 \\
    Epochs                  & 50 \\
    Learning rate           & $2 \times 10^{-5}$ \\
    Weight decay            & 0.01 \\
    LR warmup               & 10\% of steps \\
    Gradient clipping       & 1.0 \\
    Positive pair criterion & $\geq 3$ shared references \\
    \hline
  \end{tabular}
\end{table}

\subsection{Similarity computation and graph construction}

After embedding all documents with both the original and fine-tuned models, we computed pairwise cosine similarities between all papers in the corpus. These similarities serve two purposes: they define the full pairwise geometry of the embedding space and they are used to construct similarity graphs for downstream analysis and visualization.

We analyze two graph variants. The first is the complete similarity graph, in which every pair of papers is connected by an edge weighted by cosine similarity. The second is a thresholded similarity graph, in which an edge is retained only if the cosine similarity exceeds 0.94. We selected 0.94 as a sparsification threshold that preserves only strong semantic links while retaining a sufficiently large connected component for community analysis. The thresholded graph therefore highlights local neighborhood structure and community organization more clearly than the complete graph. The robustness of this threshold choice is examined in the threshold-sweep
experiments presented in Section~\ref{sec:results}.

For visualization, we compute a two-dimensional layout using the ForceAtlas2 algorithm \cite{forceatlas2}. These layouts are used only for qualitative comparison.

\subsection{Community detection and evaluation}
\label{sec:evaluation-protocol}

We apply Louvain community detection \cite{blondel2008louvain} to the thresholded graph to summarize mesoscopic structure. Modularity is reported as a graph-level measure of community organization. The resulting community labels are also used to compute embedding-space quality measures, specifically silhouette and Davies--Bouldin index, after excluding singleton communities where appropriate \cite{network_metrics}.

We report both graph-structural and embedding-quality measures. For graph structure, we evaluate isolated nodes, connected components, largest connected component, and modularity. For embedding quality, we report silhouette and Davies--Bouldin index using community assignments as labels. These values are reported for both the complete graph and the thresholded graph, since the former captures global geometry and the latter tests whether the strongest semantic links induce coherent local structure. We also compare the resulting maps qualitatively.

Two properties of this protocol should be stated explicitly, because they bound
what the resulting numbers can show. First, the community labels are not
external. For each representation, Louvain is run on that representation's own
thresholded similarity graph, and the resulting partition is then used as the
label set for silhouette and Davies--Bouldin computed in that same
representation's embedding space. Every representation is therefore scored
against a partition of its own making, and these indices measure the internal
geometric self-consistency of a representation rather than its agreement with
any external topical ground truth. Second, the complete-graph and thresholded
variants of these indices differ only in which distances enter the computation:
both use labels obtained from the thresholded graph, so neither is independent
of the sparsification step. We retain these measures because they are
informative about the compactness and separation of the resulting geometry, but
we do not treat them as evidence of semantic correctness.

A further asymmetry should be noted. Silhouette and Davies--Bouldin require a
vector space, so the bibliographic-coupling baseline, which has none, is scored
in the original SPECTER2 space. Its values therefore express the disagreement
between coupling-derived communities and SPECTER2 geometry, and are not
comparable with the other columns in the same way those columns are comparable
with each other.

\section{Results}
\label{sec:results}

\begin{table*}[htbp]
\centering
\caption{Structural and clustering-quality comparison of the bibliographic-coupling graph, the lexical TF-IDF graph, the SPECTER2 embedding graph, and the fine-tuned embedding graph. Embedding graphs are thresholded at cosine similarity $\geq 0.94$; the TF-IDF graph at cosine similarity $\geq 0.1$; the coupling graph at $\geq 2$ shared references (distinct from the $\geq 3$ criterion used to build training pairs, see Section~\ref{sec:adaptation}). `Complete graph' rows require a cosine vector space and are not applicable to the coupling graph. Silhouette and Davies--Bouldin score each graph's Louvain communities in its embedding space, the coupling communities being scored in the original SPECTER2 space; as explained in Section~\ref{sec:evaluation-protocol}, these are measures of internal self-consistency rather than of agreement with external labels. The coupling matrix covers 696 papers while the embeddings cover 704; coupling node $i$ is aligned with embedding row $i$, so the 8 papers without coupling data appear as isolated nodes in the coupling column. All values are from single runs.}
\label{tab:graph-comparison}
\begin{tabular}{lrrrr}
\hline
Metric & Coupling & TF-IDF & SPECTER2 & Fine-tuned \\
\hline
Mean similarity (complete graph)           &  \textemdash &       0.0284 &       0.9104 &       0.8663 \\
Similarity std. dev. (complete graph)      &  \textemdash &       0.0255 &       0.0259 &       0.0594 \\
Silhouette (complete graph)                &  \textemdash &       0.0098 &       0.0874 &       0.3457 \\
Davies--Bouldin (complete graph)           &  \textemdash &      10.2074 &       3.8638 &       1.7822 \\
Edges (thresholded)                        &         7560 &         3096 &        28202 &        11011 \\
Isolated nodes (thresholded)               &          331 &            7 &           68 &           16 \\
Connected components (thresholded)         &          334 &           13 &           72 &           19 \\
Largest connected component (thresholded)  &          368 &          687 &          628 &          684 \\
Modularity (thresholded)                   &       0.2334 &       0.5555 &       0.1711 &       0.5148 \\
Silhouette (thresholded)                   &      -0.1123 &       0.0214 &       0.0729 &       0.1303 \\
Davies--Bouldin (thresholded)              &       8.0413 &       5.1570 &       2.5847 &       2.3120 \\
\hline
\end{tabular}
\end{table*}

\begin{table*}[tp]
\centering
\caption{Comparison under fixed graph density. For each target density, the
globally strongest edges are retained so that all methods are evaluated
using the same edge budget (except where limited by the finite number of
bibliographic-coupling edges). This isolates differences in similarity ranking quality from differences caused solely by graph sparsity.
Silhouette and Davies--Bouldin are computed using Louvain communities in
the corresponding embedding space (bibliographic coupling is evaluated
in the original SPECTER2 space). $n=704$ papers. All values are from single runs.}
\label{tab:fixed-density}
\begin{tabular}{lrrrr}
\hline
Metric & Coupling & TF-IDF & SPECTER2 & Fine-tuned \\
\hline
\multicolumn{5}{l}{\emph{Density 1\% (target 2475 edges)}} \\
\hline
Target edges                   &         2475 &         2475 &         2475 &         2475 \\
Actual edges                   &         2475 &         2475 &         2475 &         2475 \\
Isolated nodes                 &          407 &           15 &          329 &          152 \\
Connected components           &          414 &           23 &          348 &          165 \\
Largest connected component    &          284 &          673 &          336 &          522 \\
Louvain communities ($\geq$2)  &           12 &           24 &           26 &           21 \\
Modularity                     &       0.3065 &       0.5990 &       0.3253 &       0.6309 \\
Silhouette                     &      -0.2127 &       0.0205 &       0.0783 &       0.0769 \\
Davies--Bouldin                &       6.1500 &       5.1835 &       1.7936 &       1.7966 \\
\hline
\multicolumn{5}{l}{\emph{Density 2\% (target 4949 edges)}} \\
\hline
Target edges                   &         4949 &         4949 &         4949 &         4949 \\
Actual edges                   &         4949 &         4949 &         4949 &         4949 \\
Isolated nodes                 &          361 &            2 &          249 &           70 \\
Connected components           &          364 &            4 &          263 &           75 \\
Largest connected component    &          338 &          700 &          427 &          626 \\
Louvain communities ($\geq$2)  &            9 &           14 &           23 &           12 \\
Modularity                     &       0.2575 &       0.4629 &       0.2611 &       0.5877 \\
Silhouette                     &      -0.1205 &       0.0146 &       0.0515 &       0.1057 \\
Davies--Bouldin                &       6.5138 &       6.9419 &       2.0513 &       1.9581 \\
\hline
\multicolumn{5}{l}{\emph{Density 5\% (target 12373 edges)}} \\
\hline
Target edges                   &        12373 &        12373 &        12373 &        12373 \\
Actual edges                   &        12373 &        12373 &        12373 &        12373 \\
Isolated nodes                 &          296 &            0 &          142 &           13 \\
Connected components           &          299 &            1 &          150 &           16 \\
Largest connected component    &          404 &          704 &          548 &          687 \\
Louvain communities ($\geq$2)  &            9 &           13 &           12 &            8 \\
Modularity                     &       0.2101 &       0.3246 &       0.2085 &       0.4996 \\
Silhouette                     &      -0.2035 &       0.0155 &       0.0496 &       0.1270 \\
Davies--Bouldin                &       8.2786 &       6.4443 &       2.2905 &       2.3014 \\
\hline
\multicolumn{5}{l}{\emph{Density 10\% (target 24746 edges)}} \\
\hline
Target edges                   &        24746 &        24746 &        24746 &        24746 \\
Actual edges                   &        20444 &        24746 &        24746 &        24746 \\
Isolated nodes                 &          271 &            0 &           79 &            1 \\
Connected components           &          273 &            1 &           83 &            2 \\
Largest connected component    &          431 &          704 &          617 &          703 \\
Louvain communities ($\geq$2)  &            7 &            9 &            8 &            4 \\
Modularity                     &       0.1855 &       0.2427 &       0.1772 &       0.4299 \\
Silhouette                     &      -0.1106 &       0.0135 &       0.0775 &       0.1796 \\
Davies--Bouldin                &       9.8653 &       7.5312 &       2.6341 &       2.3397 \\
\hline
\end{tabular}
\end{table*}

\begin{table*}[tp]
\centering
\caption{Comparison using symmetric $k$-nearest-neighbour graphs. Each document
is connected to its $k$ most similar neighbours and neighbourhoods are
symmetrised by union, ensuring an equal local neighbourhood budget
independently of the underlying similarity scale. This evaluates whether
the observed differences persist when local connectivity is controlled
rather than global graph density. Silhouette and Davies--Bouldin are
computed using Louvain communities in the corresponding embedding space
(bibliographic coupling is evaluated in the original SPECTER2 space).
$n=704$ papers. All values are from single runs.}
\label{tab:knn}
\begin{tabular}{lrrrr}
\hline
Metric & Coupling & TF-IDF & SPECTER2 & Fine-tuned \\
\hline
\multicolumn{5}{l}{\emph{$k = 5$}} \\
\hline
Edges                          &         1919 &         2567 &         3106 &         2740 \\
Mean degree                    &         5.45 &         7.29 &         8.82 &         7.78 \\
Isolated nodes                 &          271 &            0 &            0 &            0 \\
Connected components           &          273 &            1 &            1 &            1 \\
Largest connected component    &          431 &          704 &          704 &          704 \\
Louvain communities ($\geq$2)  &           10 &           15 &           11 &           11 \\
Modularity                     &       0.4599 &       0.5757 &       0.5257 &       0.6646 \\
Silhouette                     &      -0.1370 &       0.0109 &       0.0029 &       0.0955 \\
Davies--Bouldin                &       8.9775 &       7.1778 &       4.1227 &       2.5826 \\
\hline
\multicolumn{5}{l}{\emph{$k = 10$}} \\
\hline
Edges                          &         3583 &         4906 &         6216 &         5302 \\
Mean degree                    &        10.18 &        13.94 &        17.66 &        15.06 \\
Isolated nodes                 &          271 &            0 &            0 &            0 \\
Connected components           &          273 &            1 &            1 &            1 \\
Largest connected component    &          431 &          704 &          704 &          704 \\
Louvain communities ($\geq$2)  &            8 &           14 &            7 &            7 \\
Modularity                     &       0.3729 &       0.4608 &       0.4589 &       0.5909 \\
Silhouette                     &      -0.1108 &       0.0131 &       0.0490 &       0.1242 \\
Davies--Bouldin                &       9.5215 &       7.4740 &       4.0688 &       2.8019 \\
\hline
\multicolumn{5}{l}{\emph{$k = 15$}} \\
\hline
Edges                          &         5074 &         7206 &         9248 &         7851 \\
Mean degree                    &        14.41 &        20.47 &        26.27 &        22.30 \\
Isolated nodes                 &          271 &            0 &            0 &            0 \\
Connected components           &          273 &            1 &            1 &            1 \\
Largest connected component    &          431 &          704 &          704 &          704 \\
Louvain communities ($\geq$2)  &            8 &           11 &            7 &            8 \\
Modularity                     &       0.3276 &       0.4063 &       0.4017 &       0.5661 \\
Silhouette                     &      -0.1118 &       0.0119 &       0.0563 &       0.1253 \\
Davies--Bouldin                &      10.0396 &       8.0454 &       3.9088 &       2.6242 \\
\hline
\multicolumn{5}{l}{\emph{$k = 20$}} \\
\hline
Edges                          &         6458 &         9483 &        12275 &        10351 \\
Mean degree                    &        18.35 &        26.94 &        34.87 &        29.41 \\
Isolated nodes                 &          271 &            0 &            0 &            0 \\
Connected components           &          273 &            1 &            1 &            1 \\
Largest connected component    &          431 &          704 &          704 &          704 \\
Louvain communities ($\geq$2)  &            7 &           10 &            7 &            6 \\
Modularity                     &       0.3016 &       0.3645 &       0.3770 &       0.5413 \\
Silhouette                     &      -0.1013 &       0.0111 &       0.0669 &       0.1477 \\
Davies--Bouldin                &      11.1566 &       8.1678 &       3.7723 &       2.7085 \\
\hline
\multicolumn{5}{l}{\emph{$k = 30$}} \\
\hline
Edges                          &         8881 &        14081 &        18213 &        15160 \\
Mean degree                    &        25.23 &        40.00 &        51.74 &        43.07 \\
Isolated nodes                 &          271 &            0 &            0 &            0 \\
Connected components           &          273 &            1 &            1 &            1 \\
Largest connected component    &          431 &          704 &          704 &          704 \\
Louvain communities ($\geq$2)  &            7 &           10 &            7 &            6 \\
Modularity                     &       0.2664 &       0.3206 &       0.3378 &       0.5034 \\
Silhouette                     &      -0.1025 &       0.0127 &       0.0598 &       0.1500 \\
Davies--Bouldin                &      10.9127 &       8.0854 &       3.7620 &       2.4092 \\
\hline
\end{tabular}
\end{table*}

\begin{figure}
    \centering
    \includegraphics[width=\linewidth]{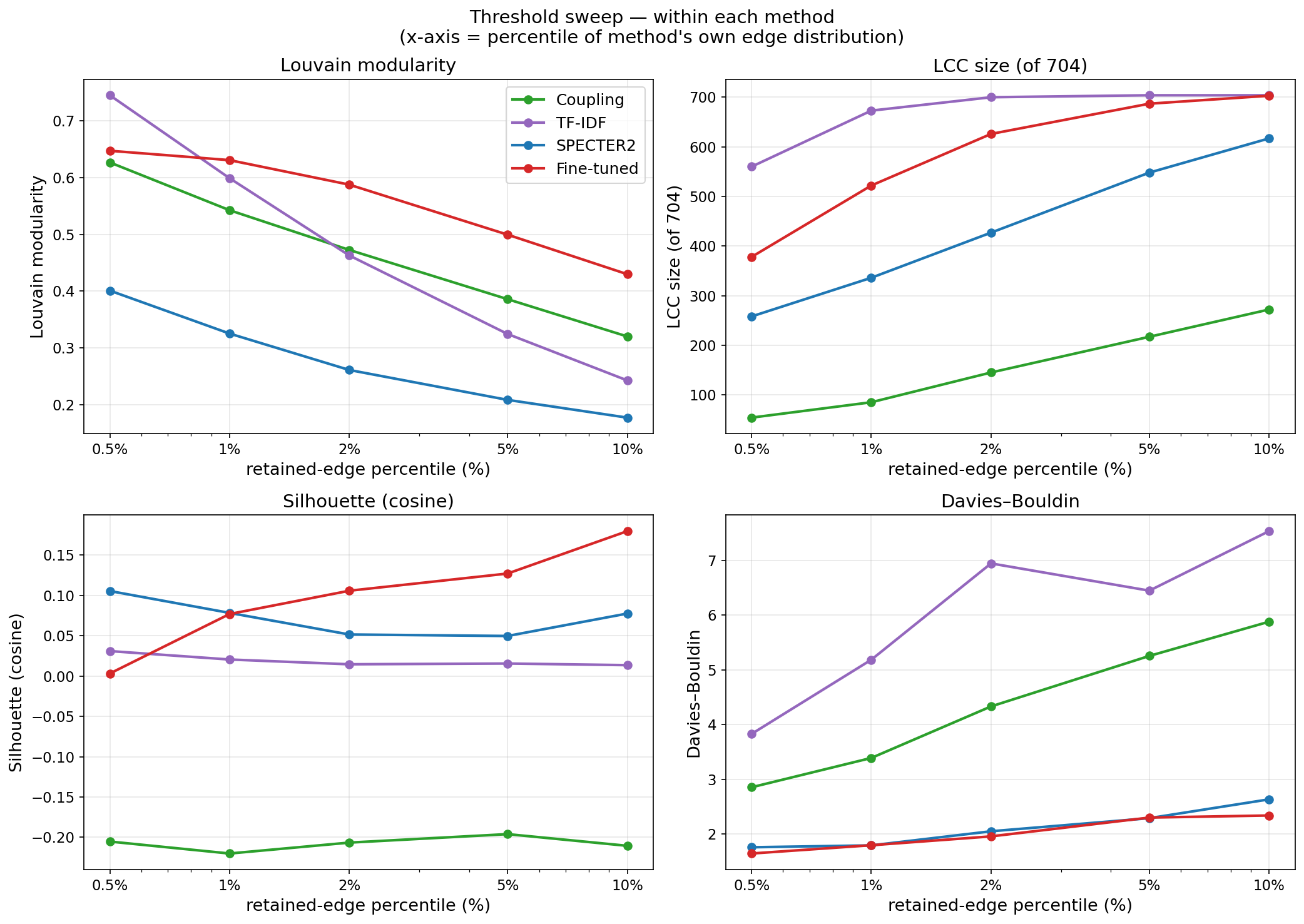}
\caption{
Threshold-sweep comparison of bibliographic coupling, TF--IDF,
original SPECTER2, and fine-tuned embeddings. Thresholds are expressed
as retained-edge percentiles within each method's own similarity
distribution rather than absolute similarity values, ensuring that
methods with different similarity scales are compared fairly. The four
panels report weighted modularity, largest connected component,
silhouette score, and Davies--Bouldin index. Across the evaluated
range, the fine-tuned embeddings provide the best balance
between community structure, graph connectivity, and clustering
quality.
}
\Description{
Four line plots compare graph quality across retained-edge
percentiles for bibliographic coupling, TF–IDF, the original
SPECTER2 embeddings, and the fine-tuned embeddings. The plots show
weighted modularity, size of the largest connected component,
silhouette score, and Davies–Bouldin index. Across almost the entire
range of graph densities, the fine-tuned embeddings achieve a better
balance between community structure, graph connectivity, and
clustering quality than the original SPECTER2 embeddings, while the
citation-based and TF–IDF baselines exhibit either fragmented graphs
or weaker semantic clustering.
}
    \label{fig:threshold-sweep}
\end{figure}

\begin{figure}[h]
    \centering
    \includegraphics[width=0.99\columnwidth]{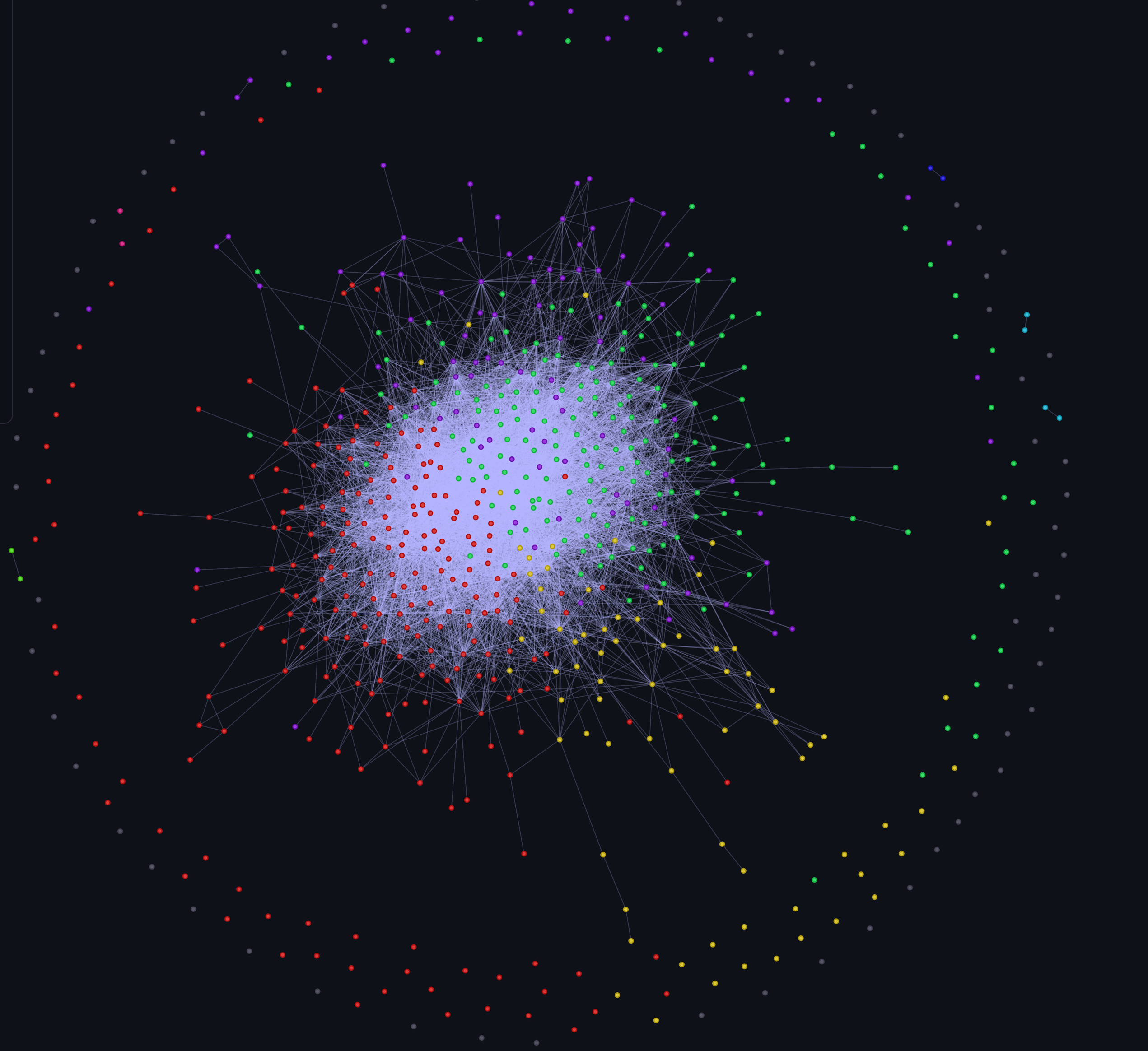}\\
    \includegraphics[width=0.99\columnwidth]{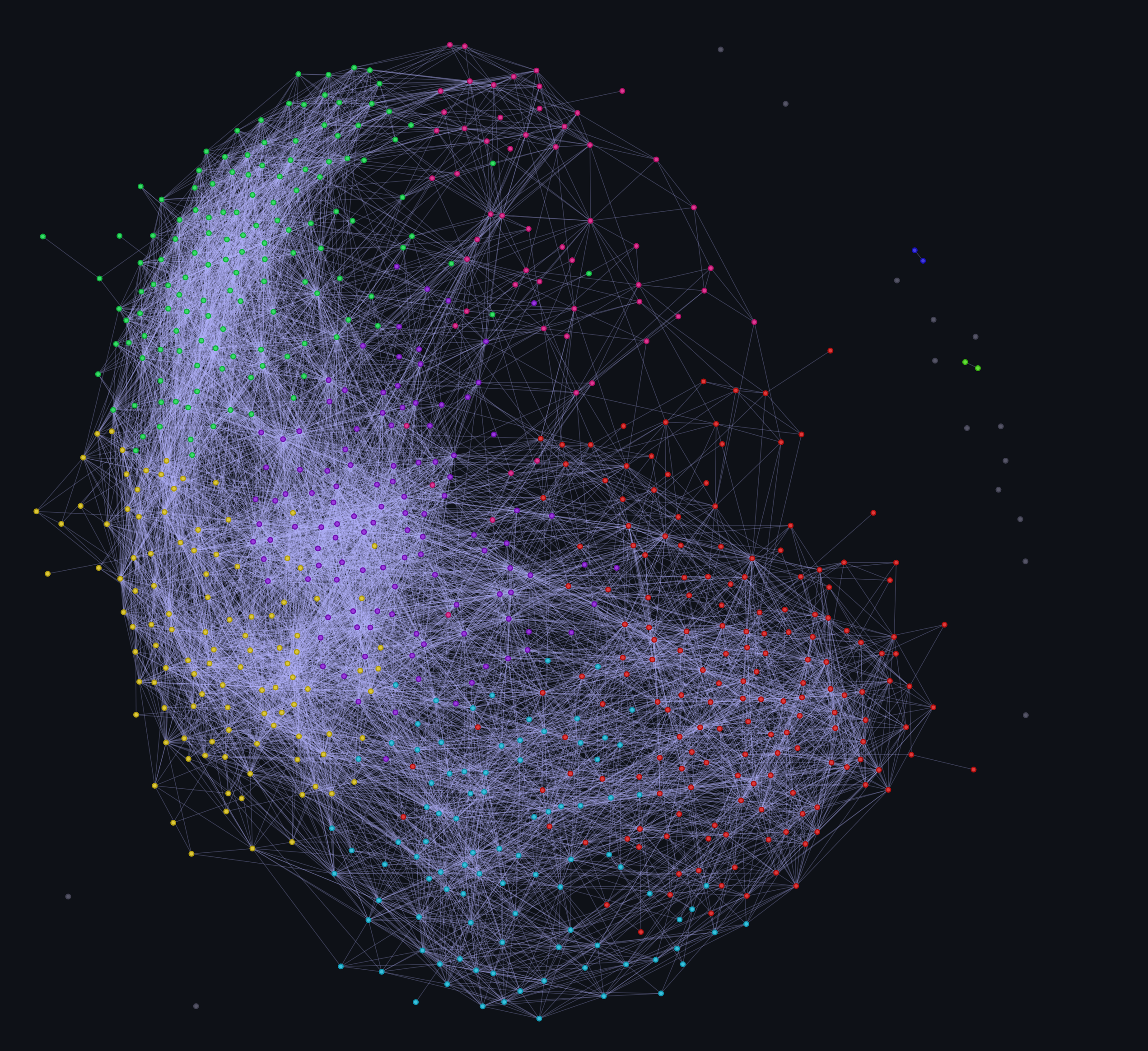}
\caption{
ForceAtlas2 layouts of literature maps constructed from the original
(top) and fine-tuned (bottom) SPECTER2 embeddings using identical
graph-construction parameters. Node colours indicate Louvain
communities. Compared with the original embedding space, the adapted
embeddings produce more clearly separated thematic regions while
preserving a dominant connected component, visually illustrating the
structural differences quantified in the graph-based
evaluation.
}
\Description{
Two ForceAtlas2 layouts of the ArtCon literature graph using identical
graph-construction parameters. The upper layout is produced from the
original SPECTER2 embeddings and forms a dense central cluster with
weakly separated communities. The lower layout uses the fine-tuned
embeddings and exhibits more clearly separated thematic regions while
preserving a large connected component. Node colours indicate
communities detected by the Louvain algorithm.
}
    \label{fig:graphs}
\end{figure}

\begin{figure}[h]
    \centering
    \includegraphics[width=0.99\columnwidth]{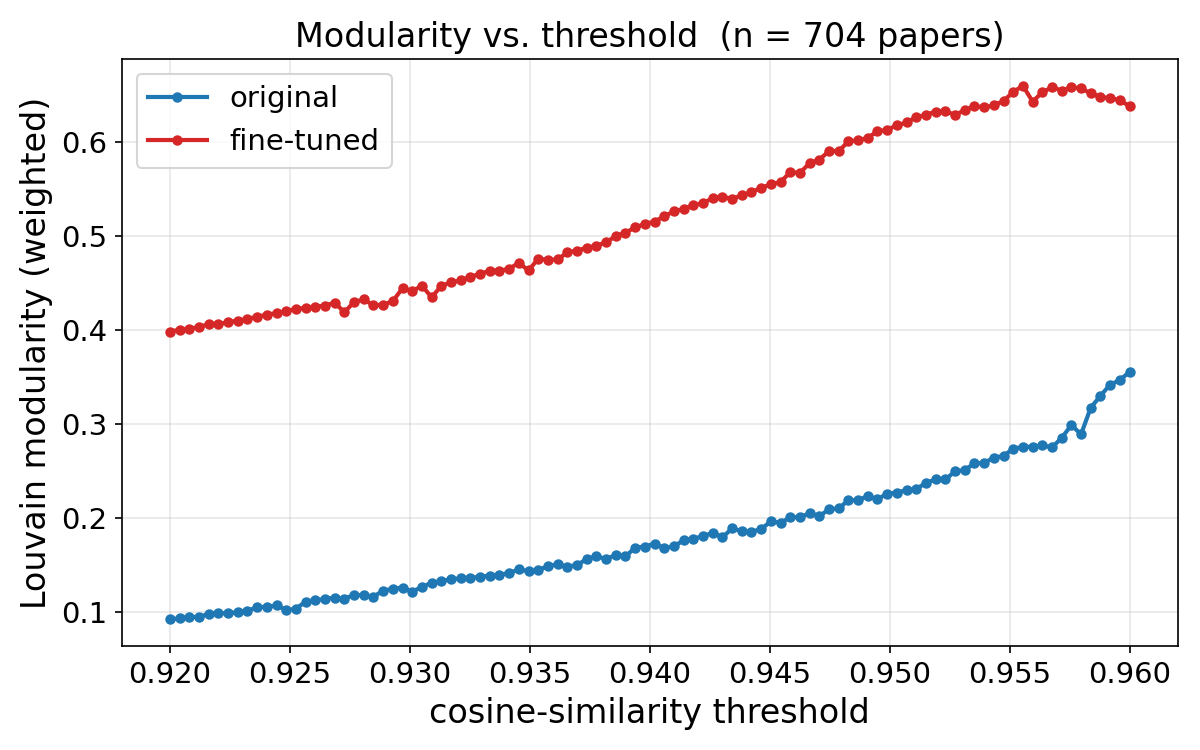}\\
    \includegraphics[width=0.99\columnwidth]{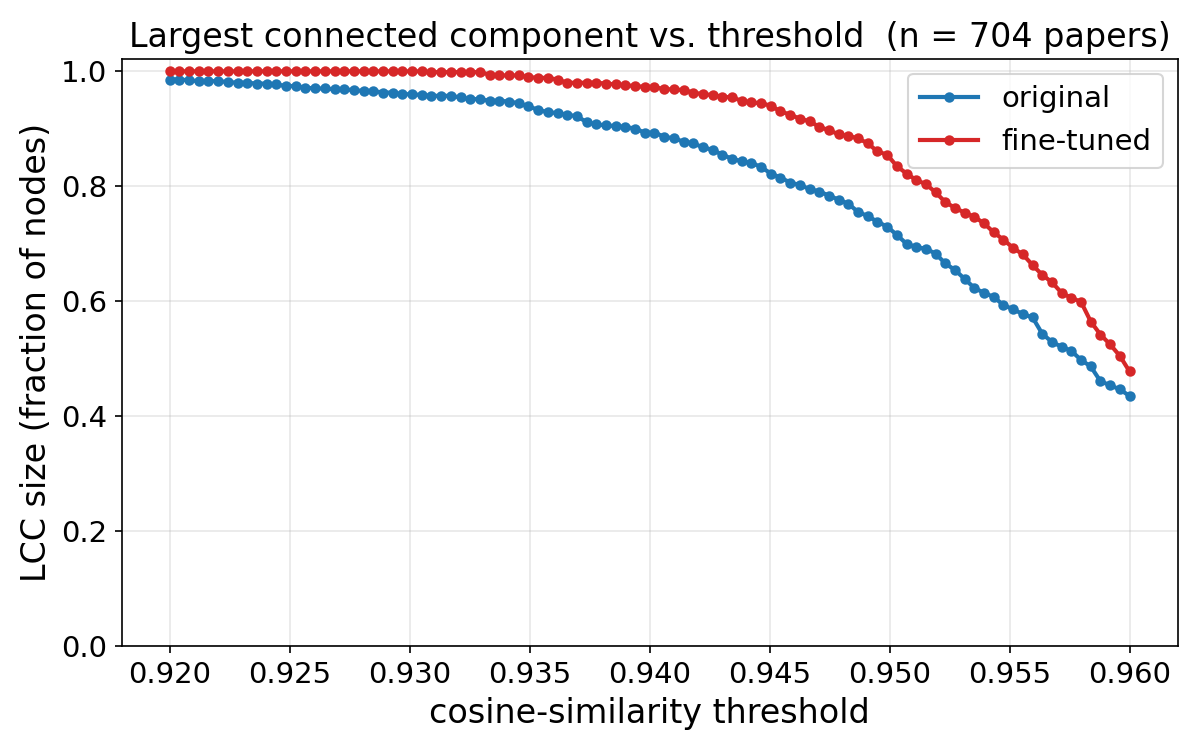}
\caption{
Sensitivity of graph structure to the cosine-similarity threshold.
Top: weighted Louvain modularity. Bottom: fraction of papers contained
in the largest connected component. The fine-tuned embeddings
consistently achieve higher modularity while preserving larger
connected components across the evaluated thresholds, indicating
that the observed structural differences are robust to graph sparsification.
}
\Description{
Two line plots compare the original and fine-tuned embeddings as the
cosine-similarity threshold increases. The upper plot shows weighted
Louvain modularity, which remains consistently higher for the
fine-tuned embeddings. The lower plot shows the fraction of papers in
the largest connected component, which also remains consistently
higher across the evaluated thresholds. Together the plots indicate
that the stronger community structure is maintained without excessive
fragmentation as graph sparsity increases.
}
    \label{fig:thresholds}
\end{figure}

\begin{table*}[t]
\centering
\caption{Illustrative changes in local semantic neighbourhoods after domain adaptation. For each anchor paper, the five nearest neighbours retrieved by the original SPECTER2 embeddings are compared with those retrieved after fine-tuning. These two anchors were selected by the authors as illustrations and were inspected without blinding to the model condition; they are not a sample, and no quantitative claim about the remaining 702 papers follows from them. Section~\ref{sec:limitations} discusses what a defensible neighbourhood evaluation would require.}
\label{tab:nn_examples}
\scriptsize
\setlength{\tabcolsep}{3pt}
\begin{tabular}{p{0.18\linewidth} p{0.38\linewidth} p{0.38\linewidth}}
\hline
Anchor & Original top-5 & Fine-tuned top-5 \\
\hline
Boltuc (2009) &
\begin{tabular}[t]{@{}l@{}}
(1) Boltuc (2012), Engineering Thesis \\
(2) Garrido-Merch\'an (2024), MC as Pseudoscience \\
(3) Susskind (2025), Conscious Machines? \\
(4) Chella \& Manzotti (2009), Manifesto for Robotics \\
(5) Arrabales (2014), Haikonen's view
\end{tabular}
&
\begin{tabular}[t]{@{}l@{}}
(1) Boltuc (2012), Engineering Thesis \\
(2) Sloman (2010), Alternative to Working on MC \\
(3) Nath (2014), Mind \& Consciousness \\
(4) Charles (2019), Discerning AC from AI \\
(5) Stuart (2007), Cognitive \& Kinaesthetic Imagination
\end{tabular}
\\
\hline
Chella \& Manzotti (2009) &
\begin{tabular}[t]{@{}l@{}}
(1) Susskind (2025), Conscious Machines? \\
(2) Myakala (2024), Critical Exploration \\
(3) Garrido-Merch\'an (2024), MC as Pseudoscience \\
(4) Aleksander (2009), Impact in Science and Engineering \\
(5) Unknown author (2003), MC
\end{tabular}
&
\begin{tabular}[t]{@{}l@{}}
(1) Kitamura (2002), Architecture for Mobile Robot \\
(2) Aleksander (2009), Impact in Science and Engineering \\
(3) Chatterjee (2012), Blueprint for AC \\
(4) Perconti \& Plebe (2025), AC \\
(5) Andreae (1987), Design of a Conscious Robot
\end{tabular}
\\
\hline
\end{tabular}
\end{table*}

We evaluate how weakly supervised adaptation changes scientific document embeddings used for exploratory literature mapping. The proposed method is compared against three baselines representing lexical similarity (TF–IDF), citation similarity (bibliographic coupling), and a general-purpose scientific embedding model (SPECTER2).
The comparison
addresses two questions. First, we investigate how the different
representations organize the global structure of a specialized
scholarly corpus. Second, we evaluate whether the observed differences
remain consistent under alternative graph-construction strategies.

Table~\ref{tab:graph-comparison} summarizes the comparison under the
standard graph construction described in Section~\ref{sec:method}.
Tables~\ref{tab:fixed-density} and~\ref{tab:knn} examine robustness
using fixed-density and symmetric $k$-nearest-neighbour graphs.

\subsection{Global organization of the representation spaces}

Table~\ref{tab:graph-comparison} compares four document
representations using thresholded similarity graphs. These
representations capture different notions of relatedness:
bibliographic coupling captures shared citation behaviour, TF--IDF
captures lexical overlap, SPECTER2 provides a general-purpose
semantic representation, and the proposed model corresponds to its
domain-adapted version.

For exploratory literature mapping, a useful representation should
produce interpretable communities while preserving enough connectivity
to support navigation across the corpus. The lexical and citation-based baselines satisfy these requirements
only partially. TF--IDF produces the highest modularity under the
thresholded setting (0.5555), ahead of the adapted embeddings (0.5148),
indicating strong lexical clustering.
However, this comes with a poor global geometry,
as reflected by its near-zero complete-graph silhouette (0.0098) and
very high Davies--Bouldin index (10.2074). Lexical
overlap is thus sufficient to recover vocabulary-based communities, but does
not by itself produce a vector space in which those communities are compact or
well separated.

Bibliographic coupling shows the opposite limitation. It captures
citation relationships directly, but produces a highly fragmented
graph, with 331 isolated nodes and 334 connected components. Its
negative silhouette score in the thresholded setting (-0.1123)
indicates that citation-derived communities do not form compact
regions in the SPECTER2 space used for its evaluation. Read with the caveat
of Section~\ref{sec:evaluation-protocol}, this says that coupling and
SPECTER2 geometry disagree substantially; it does not establish which of
the two is closer to the thematic structure of the corpus.

The original SPECTER2 embedding space is substantially more connected,
but it is also highly compressed. Pairwise cosine similarities are
concentrated around a high mean value (0.9104) with a small standard
deviation (0.0259), making it difficult to distinguish stronger
thematic relationships from weaker ones. After adaptation, the mean
similarity decreases to 0.8663 while the standard deviation more than
doubles to 0.0594. Correspondingly, the complete-graph silhouette
increases from 0.0874 to 0.3457 and the Davies--Bouldin index
decreases from 3.8638 to 1.7822.

These complete-graph indices are computed over all pairwise distances in
the embedding space and in that sense do not depend on which edges a
particular sparsification retains. They do, however, depend on the community
labels, which are obtained by running Louvain on the thresholded graph
(Section~\ref{sec:evaluation-protocol}); they are therefore neither
independent of graph construction nor independent of the representation being
evaluated. Their improvement indicates that adaptation changes the underlying
geometry rather than only the graph drawn from it, and that the new geometry
partitions more cleanly. It does not by itself indicate that the new geometry
is a more accurate account of the field.

The thresholded graph shows the same pattern. Although the adapted
representation produces fewer edges than the original SPECTER2 graph
(11\,011 versus 28\,202), it has fewer isolated nodes, fewer connected
components, a larger largest connected component, and substantially
higher modularity (0.5148 versus 0.1711). Thus, adaptation removes
many weak links while retaining stronger ones,
producing a graph that is both sparser and better connected.
These observations indicate that the adapted representation is not
simply sparser, but more selective in which pairs it ranks highly.

\subsection{Robustness to graph construction}

The preceding comparison relies on thresholded similarity graphs.
However, raw similarity values are not directly comparable across
lexical, citation-based, and embedding-based representations because
each method produces a different similarity distribution. Rather than
evaluating identical similarity thresholds, the threshold sweep uses
retained-edge percentiles computed separately for each representation.
At each point on the horizontal axis, the graph contains the same
fraction of the strongest edges available for that particular method.
This normalizes differences in similarity scale and allows the graph
structures induced by the competing representations to be compared on
equal terms.

Figure~\ref{fig:threshold-sweep} presents this sweep. Across the evaluated
range, the fine-tuned embeddings achieve the best overall balance
between community structure, graph connectivity, and clustering
quality. In particular, silhouette generally improves as progressively
weaker edges are removed, whereas TF--IDF remains close to zero despite
its high modularity. These results indicate that the differences
produced by the adaptation reflect the ordering of pairwise similarities
rather than a shift in similarity scale.
Figure~\ref{fig:threshold-sweep} further shows that only the fine-tuned embeddings improve simultaneously in modularity, clustering quality and graph connectivity across nearly the entire retained-edge range, whereas the competing representations improve at most one of these aspects.

Figure~\ref{fig:thresholds} complements this analysis by examining the
sensitivity of the two embedding models to absolute cosine thresholds.
The adapted representation consistently achieves higher modularity
while maintaining a larger largest connected component throughout the
evaluated interval. This indicates a better trade-off between community
separation and graph connectivity: the adapted graph becomes more
structured without fragmenting as quickly as the graph constructed from
the original embeddings.

Table~\ref{tab:fixed-density} provides a stricter comparison by fixing
the number of retained edges. This removes the possibility that the
observed differences are caused merely by a favourable threshold. Across all
tested densities, the fine-tuned representation has substantially
higher modularity than the original SPECTER2 embeddings. At 5\% graph
density, for example, the fine-tuned graph has far fewer isolated nodes
than SPECTER2 (13 versus 142), a much larger largest connected
component (687 versus 548), and substantially higher modularity
(0.4996 versus 0.2085). Under the same edge budget, the adapted
representation therefore allocates edges more effectively: fewer links
are spent on weak or diffuse similarities, and more of the graph
structure contributes to coherent thematic organization.
This suggests that the difference arises from a better ordering of
pairwise similarities rather than from increasing graph density.

The picture is not uniform across densities, and it is worth stating where it
is not. At the sparsest setting tested (1\% density) the fine-tuned and
original embeddings are indistinguishable on both internal clustering indices:
silhouette 0.0769 versus 0.0783 and Davies--Bouldin 1.7966 versus 1.7936, in
each case favouring the original by a margin far too small to interpret. The
Davies--Bouldin values are likewise effectively tied at 5\% density (2.3014
versus 2.2905). Since every number reported in this paper comes from a single
run, with no seed variation and no confidence intervals, differences of this
magnitude should be read as no difference. The modularity and connectivity
gaps, by contrast, are large at every density tested and are the results on
which we rely.

The fixed-density comparison also clarifies the role of TF--IDF.
TF--IDF yields very high connectivity, often producing a single
component, but its clustering-quality measures remain weak. This
suggests that lexical overlap is useful for linking papers that share
vocabulary, but less effective at producing a well-separated semantic
space. The adapted embedding model combines the advantages of both
views: it preserves high connectivity while achieving much stronger
separation of its communities than TF--IDF.

Table~\ref{tab:knn} evaluates symmetric $k$-nearest-neighbour graphs.
Unlike thresholded graphs, $k$-NN graphs normalize local connectivity
by assigning each document the same neighbourhood budget. Differences in
this setting therefore do not depend on global similarity calibration, but
reflect the composition of local neighbourhoods. Across all tested values of
$k$, the fine-tuned representation achieves the highest modularity, and it
obtains better silhouette and Davies--Bouldin scores than
the original SPECTER2 embeddings at every $k$. The adaptation therefore
changes not only global graph structure but also the local
neighbourhoods on which exploratory browsing depends.

Taken together, the threshold sweep, fixed-density comparison, and
$k$-NN comparison support the same conclusion: the structural differences are
not an artefact of a particular graph-construction procedure. The adapted
representation produces a geometry whose partitioning behaviour differs from
the original under several distinct ways of converting document similarities
into literature maps.

\subsection{Qualitative analysis}

The quantitative differences are reflected in the qualitative
organization of the corpus. Figure~\ref{fig:graphs} compares
literature maps constructed from the original and adapted SPECTER2
embeddings using the same graph-construction procedure and ForceAtlas2
layout.

The graph derived from the original SPECTER2 embeddings is dominated by
a large, densely connected central region in which community boundaries
are only weakly distinguishable. Although the Louvain algorithm
identifies several communities, their visual separation is limited,
suggesting that many semantic relationships remain insufficiently
discriminative for visual exploration.

In contrast, the graph obtained from the adapted embeddings exhibits a
more articulated global organization. Individual communities occupy
better separated regions of the layout while remaining connected
through a dominant component. This visual organization is consistent
with the quantitative differences in modularity, silhouette, and
Davies--Bouldin score. For exploratory interfaces, this is desirable because users can more
easily identify coherent thematic regions while still navigating
between related research areas.

Table~\ref{tab:nn_examples} illustrates how these structural changes
affect local neighbourhoods. For the philosophical paper by Boltuc
(2009), the original embedding retrieves a heterogeneous mixture of
philosophical, engineering, and broadly framed discussions of machine
consciousness. After adaptation, the neighbourhood shifts toward
conceptually related work in philosophy of mind and theoretical
treatments of consciousness.

A similar pattern appears for the robotics-oriented paper by Chella and
Manzotti (2009). The original neighbourhood contains several broadly
related papers together with conceptually distant publications,
whereas the adapted embedding retrieves work focused primarily on robot
architectures and conscious-system design. The cross-domain connection
between Boltuc (2009) and Chella and Manzotti (2009), present among
the original nearest neighbours, disappears after adaptation. Under a
topical-precision reading, the adapted representation distinguishes more
clearly between philosophical and engineering perspectives while preserving
thematic relationships within each area. Section~\ref{sec:limitations}
argues that this reading is not the only available one.

These two cases are illustrations chosen by the authors, not a sample, and
they were inspected with full knowledge of which model produced which column.
They are reported because they make the geometric changes concrete, not as
evidence about the corpus as a whole.

\subsection{Summary}

Across the experiments reported here, coupling-based adaptation changes the
organization of the ArtCon corpus in ways that are visible under every graph
construction we tried. Relative to the base encoder, the adapted embeddings
yield graphs with substantially stronger community structure and better
connectivity under matched edge and neighbourhood budgets, and a similarity
distribution that is far less compressed.

The comparison also shows why neither baseline is sufficient on its own.
Bibliographic coupling alone is too
fragmented to serve as a complete map of the corpus, while TF--IDF
captures lexical clusters without producing a vector space in which those
clusters are compact. Generic SPECTER2 embeddings provide a broad semantic
baseline but compress many documents into a uniformly similar region.
Using bibliographic coupling as weak supervision reshapes this
embedding space so that the resulting literature maps better balance
connectivity and community structure.

These observations are consistent with the hypothesis that bibliographic
coupling is a usable weak supervision signal for adapting scientific document
embeddings to a specialized corpus. Establishing that the reshaped map is a
more accurate account of the literature would require evaluation that does not
depend on the embedding space being evaluated, which we have not carried out
here.

\section{Discussion}

The results show that weakly supervised in-domain adaptation substantially changes the geometry of scientific document embeddings on a specialized corpus. Across thresholded, fixed-density, and symmetric $k$-nearest-neighbour graphs, the adapted model yields stronger community structure and better connectivity while the similarity distribution becomes markedly less compressed. Because the same pattern appears under three different ways of turning similarities into graphs, it reflects a change in the underlying representation rather than an artifact of a particular graph-construction strategy.

That bibliographic coupling is able to drive such a change is itself worth noting, because it is only an indirect approximation of semantic relatedness. Rather than encoding document meaning explicitly, shared references capture common intellectual context accumulated through scholarly practice. The signal is noisy and incomplete, yet roughly two million trainable parameters driven by it are enough to reorganize the space substantially. Whether a noisier but cheaper signal of this kind can approach what explicit semantic supervision would achieve is a question these experiments raise rather than answer; testing it would require a comparison against a model fine-tuned on human-annotated similarity judgments, which we have not performed.

The procedure is cheap in practice. It requires only citation metadata that are freely available through OpenAlex, a frozen encoder, and a small adapter, and it leaves downstream applications unchanged. This makes it an attractive candidate for curated scholarly collections, which rarely have relevance labels. Whether it should be adopted for that purpose depends on evidence that the present experiments do not supply.

\subsection{Threats to validity and limitations}
\label{sec:limitations}

We set out the limitations of this study in some detail, because several of them bear directly on how the results above should be read.

\paragraph{The evaluation is not independent of the representation it evaluates.}
As described in Section~\ref{sec:evaluation-protocol}, the community labels used for silhouette and Davies--Bouldin are obtained by running Louvain on each representation's own similarity graph. Contrastive training on coupled pairs pulls those pairs together, so a more compactly partitioned space is close to being what the training objective asks for. The internal indices therefore measure geometric self-consistency, and the improvement in them is to a substantial degree expected by construction. The claim these experiments support is that the adapted space partitions more cleanly; the claim that the resulting partition corresponds better to the conceptual structure of the artificial-consciousness literature is not tested anywhere in this paper. Scoring all representations against one shared external labelling, such as OpenAlex topics or a curator-assigned taxonomy, using ARI or NMI, would test it.

\paragraph{Adaptation and evaluation use the same documents.}
Positive pairs are constructed from, and the adapter is trained on, the same 704 papers on which all results are reported. There is no document-level train/test split, no evaluation on unseen papers, and no early stopping. With approximately 2M trainable parameters and 704 documents, transductive rearrangement of this particular corpus cannot be distinguished from a change that would generalize. A split of paper \emph{pairs} would not resolve this, since the same document can appear on both sides; a node-disjoint or temporal split is required. Until such an evaluation exists, the results should be read as describing the reorganization of one fixed corpus.

\paragraph{The effect has not been separated from simple geometric recalibration.}
Lower mean similarity and higher variance can also be produced by post-processing that involves no learning at all, such as mean-centering or whitening of the base embeddings. We report no such control, nor an untrained-adapter control, nor a control in which positive pairs are drawn at random with a matched degree distribution, nor a late-fusion baseline that simply combines SPECTER2 cosine similarity with coupling without any fine-tuning. Without these, the contribution of the coupling signal specifically, as opposed to any perturbation that decompresses an anisotropic space, remains undetermined.

\paragraph{No comparison with closely related citation-informed encoders.}
SciNCL \cite{ostendorff-etal-2022-neighborhood} and OpenMSD \cite{gao-etal-2024-openmsd} both learn scientific-document representations from citation relations, the latter using bibliographic coupling among its positive-pair signals. Neither is evaluated here. The comparison that would establish whether in-domain coupling-based adaptation adds anything over existing citation-aware representations is therefore missing.

\paragraph{Single corpus, single run.}
All conclusions rest on one curated interdisciplinary corpus of 704 papers. Corpora in fields with sparser or less consistent citation practices may behave differently, and bibliographic coupling may be a weaker signal there. Furthermore, every number reported is from a single run, with no random seeds recorded, no standard deviations, no significance tests and no community-stability analysis, even though training, Louvain and ForceAtlas2 are all stochastic. Small differences in the tables should not be interpreted, as noted in Section~\ref{sec:results}.

\paragraph{The qualitative evidence is illustrative only.}
Table~\ref{tab:nn_examples} reports two anchor papers, selected by the authors and inspected without blinding to the model condition. A defensible neighbourhood evaluation would sample anchors at random, present the neighbourhoods of each model in randomized order to raters blinded to their source, and report inter-rater agreement alongside per-model precision at $k$.

\paragraph{Sharper communities may have a cost.}
One change highlighted in Table~\ref{tab:nn_examples} deserves a less one-sided reading than we gave it above. The cross-domain link between the philosophical paper by Boltuc (2009) and the robotics-oriented paper by Chella and Manzotti (2009) disappears after adaptation. Under a topical-precision criterion this is an improvement. But exploratory literature mapping is valued precisely for surfacing unexpected interdisciplinary connections, and training an encoder to conform to citation structure can be expected to reinforce existing citation communities and to suppress links that cross them. The evaluation reported here cannot distinguish higher topical precision from loss of cross-domain discovery. Quantifying that trade-off, for instance as the fraction of each paper's nearest neighbours lying outside its own community before and after adaptation, is part of ongoing work.

\paragraph{Scope of the evaluation.}
We do not evaluate retrieval accuracy, because the objective of this work is exploratory mapping rather than document retrieval. We also report no runtime or memory measurements, so claims about the scalability of the procedure are outside what we have measured. Finally, the positive-pair threshold and the sparsification thresholds were not tuned, and their influence on the results is unexamined.

\section{Conclusion}

We presented a lightweight procedure for adapting scientific document embeddings to a specialized research domain, using bibliographic coupling as weak supervision and parameter-efficient LoRA fine-tuning of SPECTER2, and examined its effect on ArtCon, a curated corpus of 704 publications on artificial and machine consciousness.

On this corpus, adaptation substantially decompresses the similarity distribution and produces graphs with markedly higher Louvain modularity and better connectivity than the base encoder under matched edge budgets and matched neighbourhood budgets, consistently across thresholded, fixed-density and $k$-nearest-neighbour constructions. TF--IDF retains an advantage in thresholded modularity, and at the sparsest density tested the internal clustering indices of the adapted and original embeddings are indistinguishable.

What these experiments establish is that coupling-based adaptation changes the geometry of the embedding space in a way that makes it easier to partition into well-separated communities, and that this change survives every graph-construction choice we varied. What they do not establish is that the resulting map is a more accurate picture of the conceptual structure of the field, because every measure reported here is computed against labels derived from the representation under evaluation, and because adaptation and evaluation share the same documents.

Work in progress addresses this directly: evaluation against external topic labels, evaluation on held-out documents under a node-disjoint split, controls testing whether simple geometric post-processing of the base embeddings reproduces the same effect, comparison with existing citation-informed encoders, and replication on a second corpus from a field with different citation practices.

\section*{Data and code availability}

The ArtCon bibliography is maintained by the Institute of Philosophy of the Czech Academy of Sciences and is publicly available \cite{artcon}. Code for the adaptation and evaluation pipeline is available at \cite{github}.

\begin{acks}
  This work was partially supported by the project OP JAK: Knowledge in the Age of Distrust, CZ.02.01.01/00/23\_025/0008711, and by the LINDAT/CLARIAH-CZ Research Infrastructure supported by the Ministry of Education, Youth, and Sports of the Czech Republic (Project No. LM2023062).
\end{acks}

\bibliographystyle{ACM-Reference-Format}
\bibliography{main}


\begin{thebibliography}{30}


\ifx \showCODEN    \undefined \def \showCODEN     #1{\unskip}     \fi
\ifx \showISBNx    \undefined \def \showISBNx     #1{\unskip}     \fi
\ifx \showISBNxiii \undefined \def \showISBNxiii  #1{\unskip}     \fi
\ifx \showISSN     \undefined \def \showISSN      #1{\unskip}     \fi
\ifx \showLCCN     \undefined \def \showLCCN      #1{\unskip}     \fi
\ifx \shownote     \undefined \def \shownote      #1{#1}          \fi
\ifx \showarticletitle \undefined \def \showarticletitle #1{#1}   \fi
\ifx \showURL      \undefined \def \showURL       {\relax}        \fi
\providecommand\bibfield[2]{#2}
\providecommand\bibinfo[2]{#2}
\providecommand\natexlab[1]{#1}
\providecommand\showeprint[2][]{arXiv:#2}

\bibitem[Angelov(2020)]%
        {angelov2020top2vec}
\bibfield{author}{\bibinfo{person}{Dimo Angelov}.}
  \bibinfo{year}{2020}\natexlab{}.
\newblock \showarticletitle{Top2Vec: Distributed Representations of Topics}.
\newblock \bibinfo{journal}{\emph{CoRR}}  \bibinfo{volume}{abs/2008.09470}
  (\bibinfo{year}{2020}).
\newblock
\showeprint[arXiv]{2008.09470}
\urldef\tempurl%
\url{https://arxiv.org/abs/2008.09470}
\showURL{%
\tempurl}


\bibitem[Aria and Cuccurullo(2017)]%
        {bibliometrix}
\bibfield{author}{\bibinfo{person}{Massimo Aria} {and} \bibinfo{person}{Corrado
  Cuccurullo}.} \bibinfo{year}{2017}\natexlab{}.
\newblock \showarticletitle{bibliometrix: An {R}-tool for comprehensive science
  mapping analysis}.
\newblock \bibinfo{journal}{\emph{Journal of Informetrics}}
  \bibinfo{volume}{11}, \bibinfo{number}{4} (\bibinfo{year}{2017}),
  \bibinfo{pages}{959--975}.
\newblock
\href{https://doi.org/10.1016/j.joi.2017.08.007}{doi:\nolinkurl{10.1016/j.joi.2017.08.007}}


\bibitem[Bastian et~al\mbox{.}(2009)]%
        {bastian2009gephi}
\bibfield{author}{\bibinfo{person}{Mathieu Bastian}, \bibinfo{person}{Sebastien
  Heymann}, {and} \bibinfo{person}{Mathieu Jacomy}.}
  \bibinfo{year}{2009}\natexlab{}.
\newblock \showarticletitle{Gephi: An Open Source Software for Exploring and
  Manipulating Networks}.
\newblock \bibinfo{journal}{\emph{Proceedings of the International AAAI
  Conference on Web and Social Media}} \bibinfo{volume}{3}, \bibinfo{number}{1}
  (\bibinfo{date}{Mar.} \bibinfo{year}{2009}), \bibinfo{pages}{361--362}.
\newblock
\href{https://doi.org/10.1609/icwsm.v3i1.13937}{doi:\nolinkurl{10.1609/icwsm.v3i1.13937}}


\bibitem[Blei et~al\mbox{.}(2003)]%
        {blei2003lda}
\bibfield{author}{\bibinfo{person}{David~M. Blei}, \bibinfo{person}{Andrew~Y.
  Ng}, {and} \bibinfo{person}{Michael~I. Jordan}.}
  \bibinfo{year}{2003}\natexlab{}.
\newblock \showarticletitle{Latent dirichlet allocation}.
\newblock \bibinfo{journal}{\emph{J. Mach. Learn. Res.}} \bibinfo{volume}{3},
  \bibinfo{number}{null} (\bibinfo{date}{March} \bibinfo{year}{2003}),
  \bibinfo{pages}{993–1022}.
\newblock
\showISSN{1532-4435}


\bibitem[Blondel et~al\mbox{.}(2008)]%
        {blondel2008louvain}
\bibfield{author}{\bibinfo{person}{Vincent~D Blondel},
  \bibinfo{person}{Jean-Loup Guillaume}, \bibinfo{person}{Renaud Lambiotte},
  {and} \bibinfo{person}{Etienne Lefebvre}.} \bibinfo{year}{2008}\natexlab{}.
\newblock \showarticletitle{Fast unfolding of communities in large networks}.
\newblock \bibinfo{journal}{\emph{Journal of Statistical Mechanics: Theory and
  Experiment}} \bibinfo{volume}{2008}, \bibinfo{number}{10}
  (\bibinfo{year}{2008}), \bibinfo{pages}{P10008}.
\newblock
\href{https://doi.org/10.1088/1742-5468/2008/10/P10008}{doi:\nolinkurl{10.1088/1742-5468/2008/10/P10008}}


\bibitem[Boyack and Klavans(2010)]%
        {boyack2010coclassification}
\bibfield{author}{\bibinfo{person}{Kevin Boyack} {and} \bibinfo{person}{Richard
  Klavans}.} \bibinfo{year}{2010}\natexlab{}.
\newblock \showarticletitle{Co-Citation Analysis, Bibliographic Coupling, and
  Direct Citation: Which Citation Approach Represents the Research Front Most
  Accurately?}
\newblock \bibinfo{journal}{\emph{Journal of the American Society for
  Information Science and Technology}}  \bibinfo{volume}{61}
  (\bibinfo{date}{12} \bibinfo{year}{2010}), \bibinfo{pages}{2389--2404}.
\newblock
\href{https://doi.org/10.1002/asi.21419}{doi:\nolinkurl{10.1002/asi.21419}}


\bibitem[Campello et~al\mbox{.}(2013)]%
        {campello2013hdbscan}
\bibfield{author}{\bibinfo{person}{Ricardo J. G.~B. Campello},
  \bibinfo{person}{Davoud Moulavi}, {and} \bibinfo{person}{Joerg Sander}.}
  \bibinfo{year}{2013}\natexlab{}.
\newblock \showarticletitle{Density-Based Clustering Based on Hierarchical
  Density Estimates}. In \bibinfo{booktitle}{\emph{Advances in Knowledge
  Discovery and Data Mining}}, \bibfield{editor}{\bibinfo{person}{Jian Pei},
  \bibinfo{person}{Vincent~S. Tseng}, \bibinfo{person}{Longbing Cao},
  \bibinfo{person}{Hiroshi Motoda}, {and} \bibinfo{person}{Guandong Xu}}
  (Eds.). \bibinfo{publisher}{Springer Berlin Heidelberg},
  \bibinfo{address}{Berlin, Heidelberg}, \bibinfo{pages}{160--172}.
\newblock
\showISBNx{978-3-642-37456-2}


\bibitem[Chen(2006)]%
        {chen2006citespace}
\bibfield{author}{\bibinfo{person}{Chaomei Chen}.}
  \bibinfo{year}{2006}\natexlab{}.
\newblock \showarticletitle{CiteSpace II: Detecting and visualizing emerging
  trends and transient patterns in scientific literature}.
\newblock \bibinfo{journal}{\emph{Journal of the American Society for
  Information Science and Technology}} \bibinfo{volume}{57},
  \bibinfo{number}{3} (\bibinfo{year}{2006}), \bibinfo{pages}{359--377}.
\newblock
\showeprint{https://onlinelibrary.wiley.com/doi/pdf/10.1002/asi.20317}
\href{https://doi.org/10.1002/asi.20317}{doi:\nolinkurl{10.1002/asi.20317}}


\bibitem[Cobo et~al\mbox{.}(2012)]%
        {cobo2012scimat}
\bibfield{author}{\bibinfo{person}{Manuel Cobo}, \bibinfo{person}{A.G.
  L\'{o}pez-Herrera}, \bibinfo{person}{Enrique Herrera-Viedma}, {and}
  \bibinfo{person}{Francisco Herrera}.} \bibinfo{year}{2012}\natexlab{}.
\newblock \showarticletitle{SciMAT: A new science mapping analysis software
  tool}.
\newblock \bibinfo{journal}{\emph{Journal of the American Society for
  Information Science and Technology}}  \bibinfo{volume}{63}
  (\bibinfo{date}{08} \bibinfo{year}{2012}), \bibinfo{pages}{1609–1630}.
\newblock
\href{https://doi.org/10.1002/asi.22688}{doi:\nolinkurl{10.1002/asi.22688}}


\bibitem[Cohan et~al\mbox{.}(2020)]%
        {cohan-etal-2020-specter}
\bibfield{author}{\bibinfo{person}{Arman Cohan}, \bibinfo{person}{Sergey
  Feldman}, \bibinfo{person}{Iz Beltagy}, \bibinfo{person}{Doug Downey}, {and}
  \bibinfo{person}{Daniel Weld}.} \bibinfo{year}{2020}\natexlab{}.
\newblock \showarticletitle{{SPECTER}: Document-level Representation Learning
  using Citation-informed Transformers}. In
  \bibinfo{booktitle}{\emph{Proceedings of the 58th Annual Meeting of the
  Association for Computational Linguistics}},
  \bibfield{editor}{\bibinfo{person}{Dan Jurafsky}, \bibinfo{person}{Joyce
  Chai}, \bibinfo{person}{Natalie Schluter}, {and} \bibinfo{person}{Joel
  Tetreault}} (Eds.). \bibinfo{publisher}{Association for Computational
  Linguistics}, \bibinfo{address}{Online}, \bibinfo{pages}{2270--2282}.
\newblock
\href{https://doi.org/10.18653/v1/2020.acl-main.207}{doi:\nolinkurl{10.18653/v1/2020.acl-main.207}}


\bibitem[{Connected Papers}(2020)]%
        {connectedpapers}
\bibfield{author}{\bibinfo{person}{{Connected Papers}}.}
  \bibinfo{year}{2020}\natexlab{}.
\newblock \bibinfo{title}{Connected Papers}.
\newblock \bibinfo{howpublished}{\url{https://www.connectedpapers.com/}}.
\newblock
\shownote{Accessed: 2026-04-21}.
\newblock


\bibitem[Gao et~al\mbox{.}(2024)]%
        {gao-etal-2024-openmsd}
\bibfield{author}{\bibinfo{person}{Yang Gao}, \bibinfo{person}{Ji Ma},
  \bibinfo{person}{Ivan Korotkov}, \bibinfo{person}{Keith Hall},
  \bibinfo{person}{Dana Alon}, {and} \bibinfo{person}{Donald Metzler}.}
  \bibinfo{year}{2024}\natexlab{}.
\newblock \showarticletitle{{OpenMSD}: Towards Multilingual Scientific
  Documents Similarity Measurement}. In \bibinfo{booktitle}{\emph{Proceedings
  of the 2024 Joint International Conference on Computational Linguistics,
  Language Resources and Evaluation (LREC-COLING 2024)}},
  \bibfield{editor}{\bibinfo{person}{Nicoletta Calzolari},
  \bibinfo{person}{Min-Yen Kan}, \bibinfo{person}{Veronique Hoste},
  \bibinfo{person}{Alessandro Lenci}, \bibinfo{person}{Sakriani Sakti}, {and}
  \bibinfo{person}{Nianwen Xue}} (Eds.). \bibinfo{publisher}{ELRA and ICCL},
  \bibinfo{address}{Torino, Italia}, \bibinfo{pages}{12467--12480}.
\newblock
\urldef\tempurl%
\url{https://aclanthology.org/2024.lrec-main.1092/}
\showURL{%
\tempurl}


\bibitem[Grootendorst(2022)]%
        {grootendorst2022bertopic}
\bibfield{author}{\bibinfo{person}{Maarten Grootendorst}.}
  \bibinfo{year}{2022}\natexlab{}.
\newblock \bibinfo{title}{BERTopic: Neural topic modeling with a class-based
  TF-IDF procedure}.
\newblock
\shownote{arXiv:2203.05794}.
\newblock
\showeprint[arxiv]{2203.05794}~[cs.CL]
\urldef\tempurl%
\url{https://arxiv.org/abs/2203.05794}
\showURL{%
\tempurl}


\bibitem[Henderson et~al\mbox{.}(2017)]%
        {henderson2017efficient}
\bibfield{author}{\bibinfo{person}{Matthew Henderson}, \bibinfo{person}{Rami
  Al-Rfou}, \bibinfo{person}{Brian Strope}, \bibinfo{person}{Yun-Hsuan Sung},
  \bibinfo{person}{L{\'a}szl{\'o} Luk{\'a}cs}, \bibinfo{person}{Ruiqi Guo},
  \bibinfo{person}{Sanjiv Kumar}, \bibinfo{person}{Balint Miklos}, {and}
  \bibinfo{person}{Ray Kurzweil}.} \bibinfo{year}{2017}\natexlab{}.
\newblock \showarticletitle{Efficient Natural Language Response Suggestion for
  Smart Reply}. In \bibinfo{booktitle}{\emph{arXiv preprint arXiv:1705.00652}}.
\newblock
\urldef\tempurl%
\url{https://arxiv.org/abs/1705.00652}
\showURL{%
\tempurl}


\bibitem[Houlsby et~al\mbox{.}(2019)]%
        {houlsby2019adapters}
\bibfield{author}{\bibinfo{person}{Neil Houlsby}, \bibinfo{person}{Andrei
  Giurgiu}, \bibinfo{person}{Stanislaw Jastrzebski}, \bibinfo{person}{Bruna
  Morrone}, \bibinfo{person}{Quentin De~Laroussilhe}, \bibinfo{person}{Andrea
  Gesmundo}, \bibinfo{person}{Mona Attariyan}, {and} \bibinfo{person}{Sylvain
  Gelly}.} \bibinfo{year}{2019}\natexlab{}.
\newblock \showarticletitle{Parameter-Efficient Transfer Learning for {NLP}}.
  In \bibinfo{booktitle}{\emph{Proceedings of the 36th International Conference
  on Machine Learning}} \emph{(\bibinfo{series}{Proceedings of Machine Learning
  Research}, Vol.~\bibinfo{volume}{97})},
  \bibfield{editor}{\bibinfo{person}{Kamalika Chaudhuri} {and}
  \bibinfo{person}{Ruslan Salakhutdinov}} (Eds.). \bibinfo{publisher}{PMLR},
  \bibinfo{pages}{2790--2799}.
\newblock
\urldef\tempurl%
\url{https://proceedings.mlr.press/v97/houlsby19a.html}
\showURL{%
\tempurl}


\bibitem[Hu et~al\mbox{.}(2022)]%
        {hu2022lora}
\bibfield{author}{\bibinfo{person}{Edward~J. Hu}, \bibinfo{person}{Yelong
  Shen}, \bibinfo{person}{Phillip Wallis}, \bibinfo{person}{Zeyuan Allen-Zhu},
  \bibinfo{person}{Yuanzhi Li}, \bibinfo{person}{Shean Wang},
  \bibinfo{person}{Lu Wang}, {and} \bibinfo{person}{Weizhu Chen}.}
  \bibinfo{year}{2022}\natexlab{}.
\newblock \showarticletitle{{LoRA}: Low-Rank Adaptation of Large Language
  Models}. In \bibinfo{booktitle}{\emph{Proceedings of the 10th International
  Conference on Learning Representations (ICLR)}}.
\newblock
\showeprint[arxiv]{2106.09685}~[cs.CL]
\urldef\tempurl%
\url{https://arxiv.org/abs/2106.09685}
\showURL{%
\tempurl}


\bibitem[Jacomy et~al\mbox{.}(2014)]%
        {forceatlas2}
\bibfield{author}{\bibinfo{person}{Mathieu Jacomy}, \bibinfo{person}{Tommaso
  Venturini}, \bibinfo{person}{Sebastien Heymann}, {and}
  \bibinfo{person}{Mathieu Bastian}.} \bibinfo{year}{2014}\natexlab{}.
\newblock \showarticletitle{ForceAtlas2, a Continuous Graph Layout Algorithm
  for Handy Network Visualization Designed for the Gephi Software}.
\newblock \bibinfo{journal}{\emph{PLOS ONE}} \bibinfo{volume}{9},
  \bibinfo{number}{6} (\bibinfo{date}{06} \bibinfo{year}{2014}),
  \bibinfo{pages}{1--12}.
\newblock
\href{https://doi.org/10.1371/journal.pone.0098679}{doi:\nolinkurl{10.1371/journal.pone.0098679}}


\bibitem[Kessler(1963)]%
        {kessler1963}
\bibfield{author}{\bibinfo{person}{M.~M. Kessler}.}
  \bibinfo{year}{1963}\natexlab{}.
\newblock \showarticletitle{Bibliographic coupling between scientific papers}.
\newblock \bibinfo{journal}{\emph{American Documentation}}
  \bibinfo{volume}{14}, \bibinfo{number}{1} (\bibinfo{year}{1963}),
  \bibinfo{pages}{10--25}.
\newblock
\showeprint{https://onlinelibrary.wiley.com/doi/pdf/10.1002/asi.5090140103}
\href{https://doi.org/10.1002/asi.5090140103}{doi:\nolinkurl{10.1002/asi.5090140103}}


\bibitem[Kraker et~al\mbox{.}(2016)]%
        {kraker2016okmaps}
\bibfield{author}{\bibinfo{person}{Peter Kraker}, \bibinfo{person}{Christopher
  Kittel}, {and} \bibinfo{person}{Asura Enkhbayar}.}
  \bibinfo{year}{2016}\natexlab{}.
\newblock \showarticletitle{Open Knowledge Maps: Creating a Visual Interface to
  the World's Scientific Knowledge Based on Natural Language Processing}.
\newblock \bibinfo{journal}{\emph{027.7 Zeitschrift f\"{u}r Bibliothekskultur}}
   \bibinfo{volume}{4} (\bibinfo{date}{11} \bibinfo{year}{2016}),
  \bibinfo{pages}{98--103}.
\newblock
\href{https://doi.org/10.12685/027.7-4-2-157}{doi:\nolinkurl{10.12685/027.7-4-2-157}}


\bibitem[{Litmaps}(2024)]%
        {litmaps}
\bibfield{author}{\bibinfo{person}{{Litmaps}}.}
  \bibinfo{year}{2024}\natexlab{}.
\newblock \bibinfo{title}{Litmaps}.
\newblock \bibinfo{howpublished}{\url{https://app.litmaps.com/}}.
\newblock
\shownote{Version 2025-01-16 [Search tool]}.
\newblock


\bibitem[Marvan and Mihalik({[n.\,d.]})]%
        {artcon}
\bibfield{author}{\bibinfo{person}{Tomas Marvan} {and} \bibinfo{person}{Jakub
  Mihalik}.} \bibinfo{year}{[n.\,d.]}\natexlab{}.
\newblock \bibinfo{title}{{ArtCon}: A Bibliography of Scholarly Works on
  Artificial and Machine Consciousness}.
\newblock \bibinfo{howpublished}{Institute of Philosophy, Czech Academy of
  Sciences}.
\newblock
\urldef\tempurl%
\url{https://artcon.flu.cas.cz/}
\showURL{%
\tempurl}


\bibitem[McInnes et~al\mbox{.}(2020)]%
        {mcinnes2020umapuniformmanifoldapproximation}
\bibfield{author}{\bibinfo{person}{Leland McInnes}, \bibinfo{person}{John
  Healy}, {and} \bibinfo{person}{James Melville}.}
  \bibinfo{year}{2020}\natexlab{}.
\newblock \bibinfo{title}{UMAP: Uniform Manifold Approximation and Projection
  for Dimension Reduction}.
\newblock
\shownote{arXiv:1802.03426}.
\newblock
\showeprint[arxiv]{1802.03426}~[stat.ML]
\urldef\tempurl%
\url{https://arxiv.org/abs/1802.03426}
\showURL{%
\tempurl}


\bibitem[Newman(2018)]%
        {network_metrics}
\bibfield{author}{\bibinfo{person}{Mark Newman}.}
  \bibinfo{year}{2018}\natexlab{}.
\newblock \bibinfo{booktitle}{\emph{Networks}}.
\newblock \bibinfo{publisher}{Oxford University Press}.
\newblock
\showISBNx{0192527495, 9780192527493}


\bibitem[Ostendorff et~al\mbox{.}(2022)]%
        {ostendorff-etal-2022-neighborhood}
\bibfield{author}{\bibinfo{person}{Malte Ostendorff}, \bibinfo{person}{Nils
  Rethmeier}, \bibinfo{person}{Isabelle Augenstein}, \bibinfo{person}{Bela
  Gipp}, {and} \bibinfo{person}{Georg Rehm}.} \bibinfo{year}{2022}\natexlab{}.
\newblock \showarticletitle{Neighborhood Contrastive Learning for Scientific
  Document Representations with Citation Embeddings}. In
  \bibinfo{booktitle}{\emph{Proceedings of the 2022 Conference on Empirical
  Methods in Natural Language Processing}},
  \bibfield{editor}{\bibinfo{person}{Yoav Goldberg}, \bibinfo{person}{Zornitsa
  Kozareva}, {and} \bibinfo{person}{Yue Zhang}} (Eds.).
  \bibinfo{publisher}{Association for Computational Linguistics},
  \bibinfo{address}{Abu Dhabi, United Arab Emirates},
  \bibinfo{pages}{11670--11688}.
\newblock
\href{https://doi.org/10.18653/v1/2022.emnlp-main.802}{doi:\nolinkurl{10.18653/v1/2022.emnlp-main.802}}


\bibitem[Pfeiffer et~al\mbox{.}(2020)]%
        {pfeiffer-etal-2020-adapterhub}
\bibfield{author}{\bibinfo{person}{Jonas Pfeiffer}, \bibinfo{person}{Andreas
  R{\"u}ckl{\'e}}, \bibinfo{person}{Clifton Poth}, \bibinfo{person}{Aishwarya
  Kamath}, \bibinfo{person}{Ivan Vuli{\'c}}, \bibinfo{person}{Sebastian Ruder},
  \bibinfo{person}{Kyunghyun Cho}, {and} \bibinfo{person}{Iryna Gurevych}.}
  \bibinfo{year}{2020}\natexlab{}.
\newblock \showarticletitle{{A}dapter{H}ub: A Framework for Adapting
  Transformers}. In \bibinfo{booktitle}{\emph{Proceedings of the 2020
  Conference on Empirical Methods in Natural Language Processing: System
  Demonstrations}}, \bibfield{editor}{\bibinfo{person}{Qun Liu} {and}
  \bibinfo{person}{David Schlangen}} (Eds.). \bibinfo{publisher}{Association
  for Computational Linguistics}, \bibinfo{address}{Online},
  \bibinfo{pages}{46--54}.
\newblock
\href{https://doi.org/10.18653/v1/2020.emnlp-demos.7}{doi:\nolinkurl{10.18653/v1/2020.emnlp-demos.7}}


\bibitem[Priem et~al\mbox{.}(2022)]%
        {priem2022openalex}
\bibfield{author}{\bibinfo{person}{Jason Priem}, \bibinfo{person}{Heather
  Piwowar}, {and} \bibinfo{person}{Richard Orr}.}
  \bibinfo{year}{2022}\natexlab{}.
\newblock \bibinfo{title}{OpenAlex: A fully-open index of scholarly works,
  authors, venues, institutions, and concepts}.
\newblock
\shownote{arXiv:2205.01833}.
\newblock
\showeprint[arxiv]{2205.01833}~[cs.DL]
\urldef\tempurl%
\url{https://arxiv.org/abs/2205.01833}
\showURL{%
\tempurl}


\bibitem[{ResearchRabbit}(2021)]%
        {rabbit}
\bibfield{author}{\bibinfo{person}{{ResearchRabbit}}.}
  \bibinfo{year}{2021}\natexlab{}.
\newblock \bibinfo{title}{{ResearchRabbit}}.
\newblock \bibinfo{howpublished}{\url{https://www.researchrabbit.ai/}}.
\newblock
\shownote{Citation-based literature mapping tool.}.
\newblock


\bibitem[Singh et~al\mbox{.}(2023)]%
        {singh-etal-2023-scirepeval}
\bibfield{author}{\bibinfo{person}{Amanpreet Singh}, \bibinfo{person}{Mike
  D{'}Arcy}, \bibinfo{person}{Arman Cohan}, \bibinfo{person}{Doug Downey},
  {and} \bibinfo{person}{Sergey Feldman}.} \bibinfo{year}{2023}\natexlab{}.
\newblock \showarticletitle{{S}ci{R}ep{E}val: A Multi-Format Benchmark for
  Scientific Document Representations}. In
  \bibinfo{booktitle}{\emph{Proceedings of the 2023 Conference on Empirical
  Methods in Natural Language Processing}},
  \bibfield{editor}{\bibinfo{person}{Houda Bouamor}, \bibinfo{person}{Juan
  Pino}, {and} \bibinfo{person}{Kalika Bali}} (Eds.).
  \bibinfo{publisher}{Association for Computational Linguistics},
  \bibinfo{address}{Singapore}, \bibinfo{pages}{5548--5566}.
\newblock
\href{https://doi.org/10.18653/v1/2023.emnlp-main.338}{doi:\nolinkurl{10.18653/v1/2023.emnlp-main.338}}


\bibitem[Small(1973)]%
        {small1973}
\bibfield{author}{\bibinfo{person}{Henry Small}.}
  \bibinfo{year}{1973}\natexlab{}.
\newblock \showarticletitle{Co-citation in the scientific literature: A new
  measure of the relationship between two documents}.
\newblock \bibinfo{journal}{\emph{Journal of the American Society for
  Information Science}} \bibinfo{volume}{24}, \bibinfo{number}{4}
  (\bibinfo{year}{1973}), \bibinfo{pages}{265--269}.
\newblock
\showeprint{https://asistdl.onlinelibrary.wiley.com/doi/pdf/10.1002/asi.4630240406}
\href{https://doi.org/10.1002/asi.4630240406}{doi:\nolinkurl{10.1002/asi.4630240406}}


\bibitem[Vidnerov\'{a}(2026)]%
        {github}
\bibfield{author}{\bibinfo{person}{Petra Vidnerov\'{a}}.}
  \bibinfo{year}{2026}\natexlab{}.
\newblock \bibinfo{title}{{ArtCon}: Code for corpus analysis and embedding
  adaptation}.
\newblock
  \bibinfo{howpublished}{\url{https://github.com/PetraVidnerova/artcon}}.
\newblock
\shownote{Accessed: 2026-10-18}.
\newblock


\end{thebibliography}

\end{document}